\documentclass[%
 reprint,
superscriptaddress,
showpacs,
 amsmath,amssymb,
 aps,
floatfix,
]{revtex4-1}

\usepackage{graphicx}
\usepackage{dcolumn}
\newcolumntype{.}{D{.}{.}{-1}}
\usepackage{bm}
\usepackage{float,array,booktabs}

\begin{document}

\preprint{APS/123-QED}

\title{The \bm{$2_1^+ \to 3_1^+$} gamma width in $^{22}$Na and second class currents}
\author{S.~Triambak}
\email{striambak@uwc.ac.za}
\affiliation{Department of Physics and Astronomy, University of the Western Cape, P/B X17, Bellville 7535, South Africa}
\affiliation{iThemba LABS, P.O. Box 722, Somerset West 7129, South Africa}
\author{L.~Phuthu}
\affiliation{Department of Physics and Astronomy, University of the Western Cape, P/B X17, Bellville 7535, South Africa}
%
%
%
\author{A.~Garc\'ia}
 \affiliation{Department of Physics and Center for Experimental Nuclear Physics and Astrophysics, University of Washington, Seattle 98195, USA}
\author{G.\,C.~Harper}
\affiliation{Department of Physics and Center for Experimental Nuclear Physics and Astrophysics, University of Washington, Seattle 98195, USA}
\author{J.\,N.~Orce}
\affiliation{Department of Physics and Astronomy, University of the Western Cape, P/B X17, Bellville 7535, South Africa}
\author{D.\,A.~Short}
\affiliation{Department of Physics and Center for Experimental Nuclear Physics and Astrophysics, University of Washington, Seattle 98195, USA}
\author{S.\,P.\,R.~Steininger}
\affiliation{Department of Physics and Center for Experimental Nuclear Physics and Astrophysics, University of Washington, Seattle 98195, USA}
\author{A.~Diaz~Varela}
\affiliation{Department of Physics, University of Guelph, Guelph, Ontario N1G 2W1, Canada}
\author{R.~Dunlop}
\affiliation{Department of Physics, University of Guelph, Guelph, Ontario N1G 2W1, Canada}
\author{D.\,S.~Jamieson}
\affiliation{Department of Physics, University of Guelph, Guelph, Ontario N1G 2W1, Canada}
\author{W.\,A.~Richter}
\affiliation{Department of Physics and Astronomy, University of the Western Cape, P/B X17, Bellville 7535, South Africa}
\author{G.\,C.~Ball}
\affiliation{TRIUMF, 4004 Wesbrook Mall, Vancouver, British Columbia V6T 2A3 Canada}
\author{P.\,E.~Garrett}
\affiliation{Department of Physics, University of Guelph, Guelph, Ontario N1G 2W1, Canada}
\author{C.\,E.~Svensson}
\affiliation{Department of Physics, University of Guelph, Guelph, Ontario N1G 2W1, Canada}
%
 \author{C.~Wrede}
\affiliation{Department of Physics and Center for Experimental Nuclear Physics and Astrophysics, University of Washington, Seattle 98195, USA}
\affiliation{Department of Physics and Astronomy and National Superconducting Cyclotron Laboratory, Michigan State University, East Lansing, Michigan 48824-1321, USA}
\date{\today}

\begin{abstract}
\begin{description}
\item[Background] A previous measurement of the $\beta-\gamma$ directional coefficient in $^{22}$Na $\beta$ decay was used 
to extract recoil-order form factors. The data indicate the requirement of a significant induced-tensor matrix element for the decay.
This conclusion
largely relies on a Standard-Model-allowed weak magnetism form factor 
which was determined using an unpublished value of the analog $2_1^+\to3_1^+$ $\gamma$ branch in $^{22}$Na, with the further assumption that the transition is dominated by its isovector $M1$ component. 
\item[Purpose]To determine the $2_1^+\to3_1^+$ width in $^{22}$Na in order to obtain an independent measurement of the weak magnetism form factor for the $\beta$ decay.  
\item[Methods]A $^{21}{\rm Ne}(p,\gamma)$ resonance reaction on an implanted target was used to produce the first $2^+$ state in $^{22}$Na at $E_x = 1952$~keV. Deexcitation $\gamma$-rays were registered with two 100\% relative efficiency high purity germanium detectors. 
\item[Results] We obtain for the first time an unambiguous determination of the $2_1^+ \to 3_1^+$ branch in $^{22}$Na to be $0.45(8)\%$.   
\item[Conclusions]Using the Conserved Vector Current (CVC) hypothesis, our branch 
determines the weak magnetism form factor for $^{22}$Na $\beta$ decay to be $|b/Ac_1| = 8.7(1.1)$. Together with the $\beta-\gamma$ angular correlation coefficient, we obtain a large induced-tensor form factor for the decay that continues to disagree with theoretical predictions. Two plausible explanations are suggested. 
\end{description}
\end{abstract}

\maketitle
\section{Introduction}
Due to the composite nature of nucleons and the presence of the strong force, the hadronic part of the weak current is known to include momentum-dependent form factors. In particular, the matrix element of the \mbox{$n \to p$} current is made of vector and axial-vector components~\cite{Grenacs:85}
\begin{align}
 \langle p | V_\mu | n \rangle & = {\bar u}(p_1)\Big[f_V \gamma_\mu + f_M \frac{\sigma_{\mu\nu} q_\nu}{2m_p} \nonumber \\
&+i f_S q_\mu \frac{m_p + m_n}{m_\pi^2}\Big]u(p_2) 
 \label{vector}
 \end{align}
%
 \begin{align}
  \langle p | A_\mu | n \rangle & = {\bar u}(p_1)\Big[f_A \gamma_\mu \gamma_5 - f_T \frac{\sigma_{\mu\nu} \gamma_5 q_\nu}{3m_p} \nonumber \\
 &-i f_P q_\mu \gamma_5 \frac{m_p + m_n}{m_\pi^2}\Big]u(p_2)~, 
 \label{axial}
  \end{align}
 where the $u$'s are Dirac spinors, $q_\mu=(p_1-p_2)_\mu$ is the 4-momentum transfer, $f_{V,A}$ are leading-order vector and axial-vector form factors, and $f_{M,S,T,P}$ are the weak magnetism, induced-scalar, induced-tensor and pseudoscalar form factors, respectively. 
 The higher-order momentum-transfer dependent contributions are often called ``recoil-order corrections'', that are either allowed or excluded in the Standard Model based on certain symmetry properties. Weinberg~\cite{Weinberg:58} classified the weak interaction currents to be first or second-class depending on their transformation under the $G$-parity operation
 \begin{equation}
  G = Ce^{i\pi T_{2}}~,
 \end{equation}  
which is the product of the charge-conjugation operator $C$ and a rotation by 180$^\circ$ about the second axis in isospin space. Following this definition, on comparison with the others, the induced-scalar and tensor-currents 
have opposite transformation properties with respect to $G$-parity and are classified as second-class. 
In the limit of perfect isospin symmetry, second-class currents (SCCs) are forbidden in the Standard Model~\cite{Weinberg:58}. In this context, nuclear $\beta$ decay studies have played an important role in searches for SCCs~\cite{Wilkinson:00}. 
In the description of nuclear $\beta$ decays using the \textit{elementary particle approach}~\cite{Holstein:RMP}, the decay matrix element is characterized in terms of similar form factors for nuclei, so that the leading-order $\gamma_\mu$ and $\gamma_\mu \gamma_5$ terms reduce to the Fermi and Gamow-Teller operators in the non-relativistic limit. 
However, disentangling a definitive Standard-Model-forbidden second-class signal from Standard-Model-allowed effects in nuclei is challenging. 
This is because induced first-class nuclear form factors mimic second-class terms~\cite{Holstein:RMP}, in addition to the up-down quark mass difference which is known to allow a small SCC~\cite{HolsteinBook,donoghue}. 
The latter isospin-violating second-class contribution is expected to be orders of magnitude smaller than the former~\cite{donoghue,Wilkinson:00}, well beyond current experimental sensitivity. Nevertheless, an accurate understanding of first-class form factors is imperative for searches of SCCs in nuclei.
Alternative searches for SCCs in $\tau$ decays have recently regained attention due to the comparatively larger momentum transfer and the absence of nuclear structure effects in these decays~\cite{PhysRevD.94.034008,PhysRevD.86.037302,babar:09,babar:11}.
\\
\\ 
The recoil-order form factors can be experimentally extracted from nuclear $\beta$ decays using angular correlation measurements~\cite{Holstein:RMP,McKeown:80}. Furthermore, if such studies are extended to mirror nuclei (such as $^{12}$B and $^{12}$N in the $A = 12$ triplet),
second-class contributions can be isolated from induced first-class terms~\cite{HolsteinBook,Calaprice:77}.
Some evidence for SCCs, well beyond Standard-Model-allowed contributions, were claimed to have been observed in the 1970's~\cite{sugimoto:75, Calaprice:75}, but were dispelled subsequently. Refs.~\cite{minami:01,minami:11} present more recent examples of state-of-the-art experiments that have yielded the best limits on second class currents from nuclear $\beta$ decays so far.
\\
\\
In this paper we discuss the particular case of $^{22}{\rm Na}$ $\beta$ decay, which provides an opportunity to probe for SCCs due to a suppression of the Gamow-Teller matrix element ($\log ft \sim 7.4$). On expanding the Gamow-Teller term~\cite{Calaprice:76,Calaprice:77} so that it includes a second-order momentum-dependent factor 
\begin{equation}
c(q^2)\equiv c_1 + c_2 q^2 + ...~, 
\end{equation}
the leading axial-vector form factor $c_1$ can be obtained from the average of several precisely measured corrected ${\cal F}t$ values for superallowed Fermi decays~\cite{TH:15} and the $ft$ value of $^{22}$Na $\beta$ decay, so that~\cite{imix}
\begin{equation}
\label{gt}
c_1 \simeq \left(\frac{2{\cal F}t^{Fermi}}{ft(^{22}{\rm Na})}\right)^{1/2} \simeq 0.0153~.
\end{equation}
Firestone, McHarris and Holstein~\cite{Firestone:78} performed shell model calculations of recoil-order form factors for $^{22}$Na $\beta$ decay using the impulse approximation and the wavefunctions described in Ref.~\cite{Calaprice:77}. The calculations, listed in Table~\ref{tab:calcs}, yielded higher-order corrections relative to the leading Gamow-Teller term $c_1$~\cite{calcs}.  
\begin{table}[b]
\caption{\label{tab:calcs}%
Calculations of higher-order form factors for $^{22}$Na beta decay from Ref.~\cite{Firestone:78}. $A$ is nucleon number and $R$ is nuclear radius.
}
\begin{ruledtabular}
\begin{tabular}{lc}
\multicolumn{1}{c}{Form factor}&\multicolumn{1}{l}{Calculated value}\\
\colrule\\[-0.9em]
Weak magnetism $b/Ac_1$ & -19\\
Second-order axial vector $c_2/c_1R^2$&-0.37\\
First-class induced tensor $d/Ac_1$ & -3.2\\
\end{tabular}
\end{ruledtabular}
\end{table}
In light of these calculations, the currently available data present some contradictions if one considers previous measurements of the electron-capture to positron decay branching ratio~\cite{Firestone:78} and  
the most recent measurement of the $\beta-\gamma$ correlation in $^{22}$Na decay with the Gammasphere array~\cite{Bowers:99}. The authors of Ref.~\cite{Bowers:99} used the measured $\beta-\gamma$ directional coefficient $A_{22} = 5.3(2.5)\times10^{-4}$ to extract the induced-tensor form factor using the parameterization~\cite{Firestone:78,Bowers:99}
\begin{equation}
\label{eq:tensor}
\frac{d}{Ac_1} = \frac{1}{4.4}\left[A_{22}10^5+0.6 \frac{c_2}{c_1R^2}\right] - \frac{b}{Ac_1}~,  
\end{equation}
which yielded $d/Ac_1 = 26(7)$, in strong disagreement with theoretical predictions (Table~\ref{tab:calcs}). 
%
\begin{figure}[t]
\includegraphics[scale=0.35]{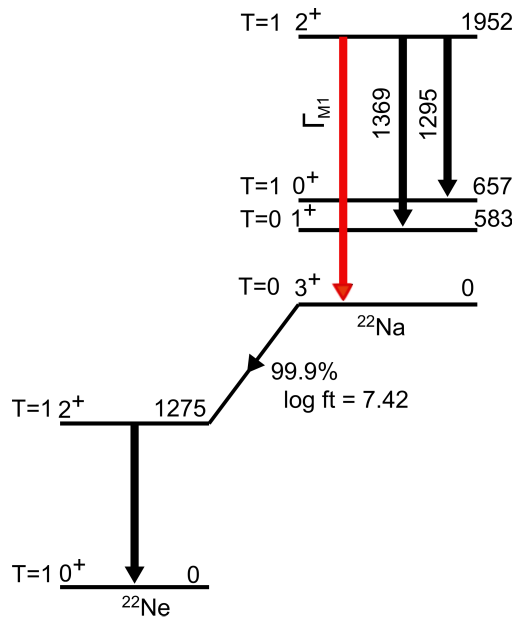}
\caption{\label{fig:22nadecays}(Color online) Decay scheme of $^{22}$Na shown with the transitions of interest. Energies are in keV.}
\end{figure}
The above conclusion was based on an unpublished determination of the weak magnetism form factor $|b/Ac_1| = 14(4)$~\cite{firestone_report} and the assumption that $b$ and $c_1$ have opposite signs, with $c_2$ being a small contribution.    
\\
\\
The anomalous induced-tensor term mentioned above calls into question the weak magnetism form factor for the decay which is not on a secure footing. This form factor
was determined using the analog $2^+ \rightarrow 3^+$ electromagnetic transition in $^{22}{\rm Na}$ (shown in Fig.~\ref{fig:22nadecays}) and the Conserved Vector Current (CVC) hypothesis~\cite{Gell-mann:58}, such that

%
%
\begin{equation}
\label{eq:weak}
b = \kappa\left(\frac{\Gamma_{M1}\cdot 6 M^2}{\alpha {E_{\gamma}}^3}\right)^{1/2}~,
\end{equation}
where $\Gamma_{M1}$ and $E_{\gamma}$ are the isovector $M1$ width and photon energy of the analog $\gamma$ transition, $M$ is the average of the parent and daughter nuclear masses, $\kappa$ is a constant~\cite{wig_eck} and $\alpha$ is the fine-structure constant.
The two experimental observables that went into determining $\Gamma_{M1}$ (and therefore $b$) in the above were the lifetime of the $E_x = 1952$~keV analog state~\cite{Bister:78} and the $1952 \to 0$~keV branch, whose value has so far only been published in a laboratory report~\cite{firestone_report} to be $0.61(24)\%$. It is thus evident that a remeasurement of this branch is an essential step
in examining the origin of the large tensor term reported in Ref.~\cite{Bowers:99}. 
\\
\\
In this paper we report the first conclusive determination of 
the aforementioned $2_1^+ \to 3_1^+$ $\gamma$ 
branch 
in $^{22}$Na to address the above issue. Two well known $^{21}{\rm Ne}(p,\gamma)$ resonances, at proton energies \mbox{$E_p = 908~\textrm{and}~1113$~keV}, were used to produce high-lying states in $^{22}$Na (at excitation energies of $7.6$~MeV and $7.8$~MeV respectively), both of which are known to predominantly feed the first $2^{+}$ state of interest~\cite{Anttila:70,Berg:77}.
\section{Experimental details}
\begin{figure}[t]
\includegraphics[scale=0.23]{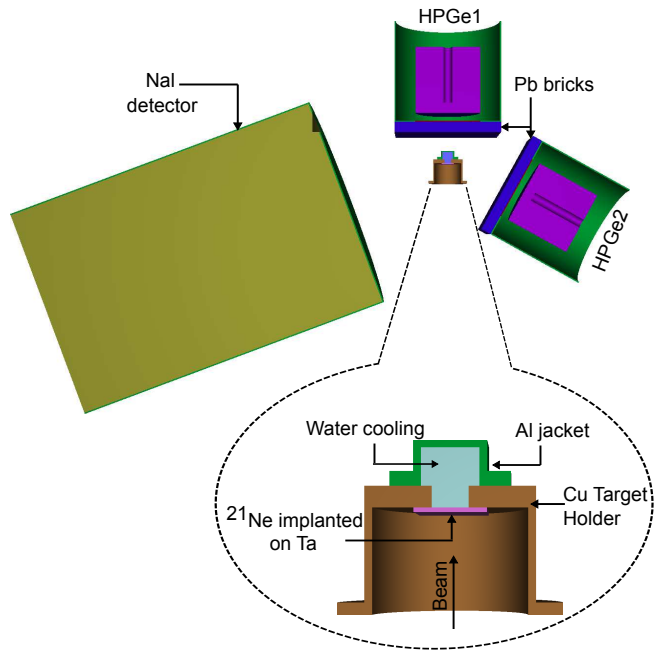}
\caption{\label{fig:setup}(Color online) Detector geometry used for this measurement. The target to detector distance for HPGe1 and HPGe2 is 4.5~cm and 8.4~cm respectively. HPGe2 is oriented at 119$^\circ$ to the beam axis. The model shown in this picture was used for the simulations mentioned in the text.}
\end{figure}
\subsection{Target preparation}
The 
targets were produced at the Center for Experimental Nuclear Physics and Astrophysics (CENPA), at the University of Washington, by implanting a 30~keV, 50~pnA $^{21}$Ne$^{++}$ beam from a modified Direct Extraction Ion Source (DEIS) into a 
1-mm-thick high-purity Tantalum backing. The beam was rastered using magnetic steerers to produce targets of thickness $\approx$~13~$\mu g$/cm$^2$ over a uniform implantation region of diameter~0.8~cm. 
\subsection{Apparatus}
The CENPA FN tandem accelerator was used as a single-ended machine with a positive (RF) ion-source placed at the terminal to produce a high intensity~$\approx~30~\mu$A~proton beam for the reaction. Target deterioration was minimized by direct water cooling on the backing and a rastering of the proton beam over the implantation region
to minimize local heating at the beam spot. 
Three detectors, placed as shown in Fig.~\ref{fig:setup}, were used to register the $\gamma$ rays emitted from the reaction. One large $10''\times15''$ NaI detector was used to collect NaI-HPGe coincidences for cross-check purposes,
while two 100$\%$ relative efficiency N-type CANBERRA HPGe detectors were used to collect the required spectra. The latter were shielded with 2.54-cm-thick lead bricks on the front to ensure negligible summing with 583~keV gamma rays from the 1952~$\to$~583~$\to$~0~keV cascade. The detector signals were digitized using  a ORTEC 413A ADC with a fast FERAbus read-out on a CAMAC crate and stored in time-stamped event mode using a java based data acquisition system~\cite{JAM}. A $^{60}$Co source of activity 970(29)~Bq~\cite{source} and a locally produced $^{56}$Co source were placed at the beam spot and used for calibration purposes.
\section{Data analysis}
\subsection{Characterization of spectra}
\begin{figure}[t]
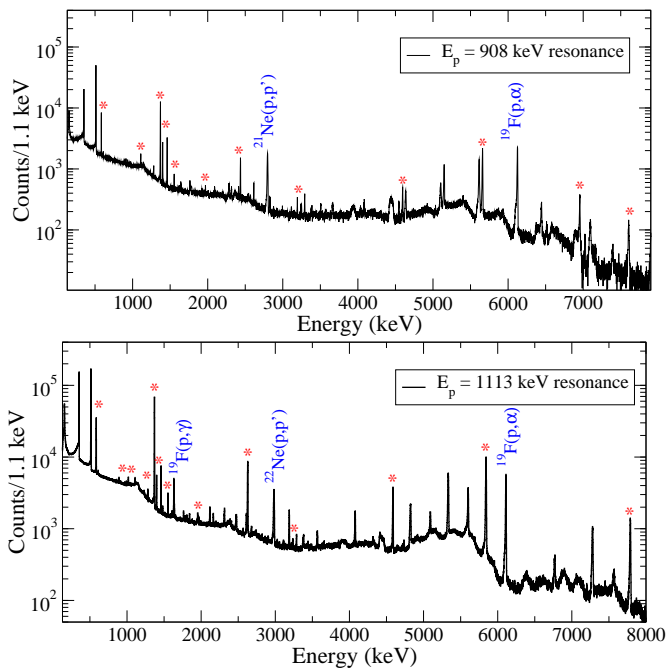

\includegraphics[scale=0.36]{21Nepg_908_singles.eps}\\
\includegraphics[scale=0.36]{21Nepg_1113_singles.eps}
\caption{\label{fig:spectrum}(Color online) Singles $^{21}{\rm Ne}(p,\gamma)$ spectra from HPGe1 for both resonances. The relevant $^{22}$Na $\gamma$-rays produced from the resonances are marked with asterisks. A few contaminant lines are present and are discussed in the text. 
}
\end{figure}

Sample $^{21}{\rm Ne}(p,\gamma)$ spectra from both resonances are shown in Fig.~\ref{fig:spectrum}. The main contaminant peaks in these spectra (other than room background) arise from $^{19}$F and $^{22}$Ne impurities in the target. The $^{19}$F contamination is commonly observed while using tantalum backings~\cite{Kontos:12} and is characterized by 
an intense 6129 keV peak from resonant $^{19}$F($p,\alpha \gamma$) reactions. The $^{22}$Ne impurity is atypical, but not unexpected. The origin of this contamination is most likely due to tails in the momentum distribution of the mass-separated ions during the implantation process. It is apparent that the lower energy resonance rendered a cleaner data set for analysis, with the contaminant peaks from $^{22}$Ne being virtually non-existent in the spectrum. The impact of the contaminants in extracting the final result is discussed in Section~\ref{results}. 

\subsection{Efficiency calibration}
\begin{figure}[t]
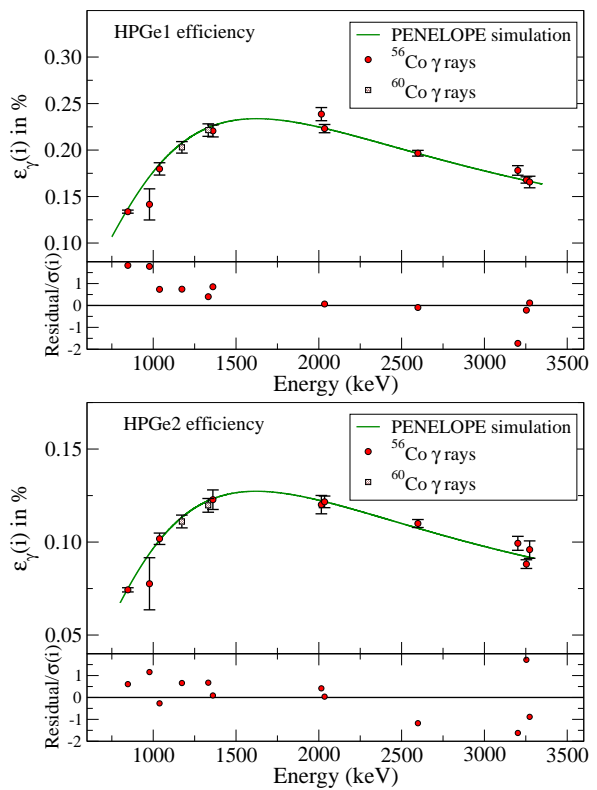

\includegraphics[scale=0.36]{sim_compare_det1_paper.eps}\\
\includegraphics[scale=0.36]{sim_compare_det2_paper.eps}
\caption{\label{fig:efficiency}(Color online) Comparison of simulated efficiencies with the experimentally determined values for both detectors. $10^6$ primary showers were used in the simulations at each \mbox{$\gamma$-ray} energy.}
\end{figure}
Since the aim of our experiment was to obtain relative intensities, our final answer is independent of the data acquisition deadtime. However, dead time corrections had to be performed for an absolute efficiency calibration of the HPGe detectors. For these corrections, the calibration sources were independently placed at the beam spot and a Berkeley Nucleonics high-precision pulser was used to send 100~Hz signals to a scalar unit on the CAMAC crate and the `test' preamplifier input of HPGe1 simutaneously. The fraction of counts lost due to the dead time in each run was determined using the ratio of the pulser peak area in the spectrum to the scaled pulser counts. These losses were found to be of order $\sim$~0.1\%. 
Dead-time-corrected peak areas were used to calculate absolute efficiencies for the germanium detectors at $\gamma$-ray energies of 1173 and 1332~keV. These values were finally used to normalize relative efficiency curves obtained from the $^{56}$Co source up to $\sim3.2$~MeV, as shown in Fig.~\ref{fig:efficiency}. 
\\
\\
Once the absolute~$\gamma$ detection efficiencies were determined over the energy range of interest, we 
obtained simulated efficiencies using the PENELOPE radiation transport code~\cite{penelope}. The model used in the simulations is shown in Fig.~\ref{fig:setup}. In the simulations monoenergetic $\gamma$ rays were emitted isotropically, originating at the beam spot on the tantalum foil shown in Fig.~\ref{fig:setup}. Several simulations were performed at different energies (846~$\le E_{\gamma} \le$~3273 keV) corresponding to the most intense peaks from the calibration sources. The events registered by the detectors were binned and used to calculate photopeak efficiencies. As shown in Fig.~\ref{fig:efficiency} the simulations agree well with the measurements.
%
\begin{table}[t]
\caption{\label{tab:effi}%
Simulated $\gamma$-ray detection efficiencies for the germanium detectors. The distributed-source-simulation had water cooling incorporated in the model. Only statistical uncertainties are shown. 
These results are for an isotropic distribution of $4 \times 10^6$ $\gamma$ rays.
}
\begin{ruledtabular}
\begin{tabular}{ccccc}
&\multicolumn{2}{c}{HPGe1 ($0^\circ$)}&\multicolumn{2}{c}{HPGe2 ($119^\circ$)}\\
$E_\gamma$&\multicolumn{2}{c}{$\epsilon_{\gamma}~(\%)$}&\multicolumn{2}{c}{$\epsilon_{\gamma}~(\%)$}\\
(keV)&\textrm{Point}&
\textrm{Distributed}&Point&Distributed\\
\colrule\\[-0.9em]
1295&0.219(2) & 0.187(2) & 0.121(2) & 0.120(2)\\
1369&0.222(2) & 0.198(2) & 0.121(2) & 0.123(2)\\
1952&0.223(2) & 0.199(2) & 0.124(2) & 0.122(2)\\
\end{tabular}
\end{ruledtabular}
\end{table}
Similar simulations were performed to obtain absolute efficiencies for the three $\gamma$ rays of interest from $^{21}{\rm Ne}(p,\gamma)$, with two important differences
\begin{enumerate}
 \item The origin of the photons was now randomly distributed on the surface of the tantalum due to the size of the implantation region and the rastering of the proton beam~\cite{comment}. 
 \item The $\gamma$ rays detected in HPGe1 were further attenuated by the water cooling on the back of the target.
\end{enumerate}

Table~\ref{tab:effi} compares the simulated efficiencies for both detectors for a point source (with no water cooling) to a distributed source (with water cooling).
It is apparent that the photo peak efficiency of HPGe1 is modified significantly on adding the water cooling and source distribution to the simulations. It should also be noted that the above efficiencies were determined assuming that the close-packed geometry of the detectors washed out angular distribution effects due to different multipolarities of the transitions. This assumption was validated by further simulations with dipolar and quadrupolar distributions for the photons, where no statistically significant deviations were observed. 

\section{Systematic effects}\label{results}
\begin{figure}[t]
\includegraphics[scale=0.34]{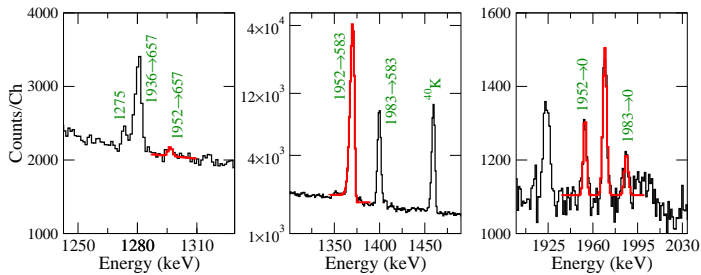}
\caption{\label{fig:fits}(Color online) HPGe1 spectrum highlighting the three $\gamma$~rays of interest (amongst others) from the 908-keV resonance and their fits.}
\end{figure}

At both resonances we unambiguously identify the three $\gamma$~rays with energies 1295, 1369 and 1952~keV (\textit{c.f.}~Fig.~\ref{fig:fits}) following the deexcitation of the 1952-keV state. These data were fit using standard functions~\cite{Triambak:06} to obtain peak areas, which were subsequently used to obtain the branching ratios.\\
\\
\subsection{$^{22}$Ne contamination}
\begin{figure}[t]
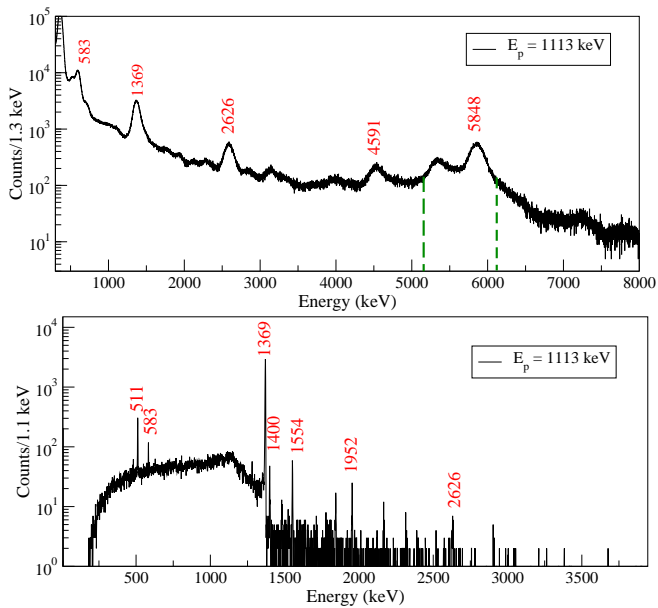

\includegraphics[scale=0.36]{nai_spec.eps}\\
\includegraphics[scale=0.36]{get1_coinc_nai_tac.eps}
\caption{\label{fig:coincs}(Color online) Top panel: Singles NaI spectrum. BottomPanel: Coincidence spectrum for HPGe1 generated by gating on the 5848~keV peak in the NaI detector as shown. The relevant peaks from $^{21}{\rm Ne}(p,\gamma)$ are labeled.}
\end{figure}

We highlight one important difference between the spectra obtained from the two resonances.  Unlike the data shown in Fig.~\ref{fig:fits}, the fits to the 1952~keV peak 
from the 1113-keV resonance yield unusually large peak widths, with FWHM's of 6.4(4)~keV for HPGe1 and 5.8(4)~keV for HPGe2.  
We conjecture that this is due to $^{22}$Ne contamination in the target. 
It is highly likely that at higher proton energies the $J^\pi = 1/2^+$,~$E_x = 2391$~keV state in $^{23}$Na~\cite{Bakkum:89} is produced in some amount via the $^{22}{\rm Ne}(p,\gamma)$ reaction. This state decays via a 2391~$\to$~440~keV transition emitting a contaminant $\gamma$~ray of energy 1951~keV. \\
\begin{figure}[t]
\includegraphics[scale=0.36]{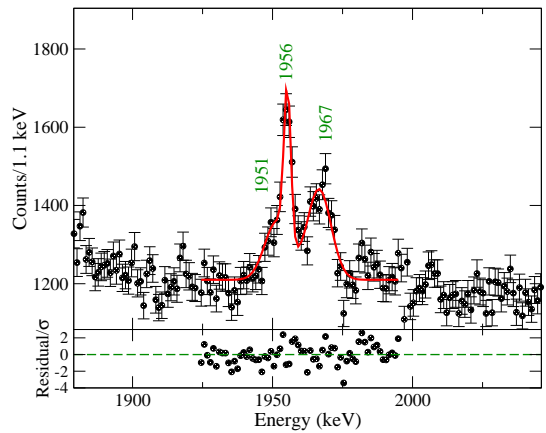}
\caption{\label{fig:fit_to_1952}(Color online) Fit to the 1952~keV peak from HPGe1 for the 1113~keV resonance. In the fit we kept the centroid of the 1951~keV peak from $^{22}{\rm Ne}(p,\gamma)$ and the width of the 1956~keV peak fixed based on the values listed in Table~\ref{tab:doppler}. The broad 1967~keV line is an escape peak of a 2988-keV $\gamma$~ray from $^{22}{\rm Ne}(p,p')$.}
\end{figure}
%
\begin{table}[t]
\caption{\label{tab:doppler}%
Simulated Doppler effects on the 1952~keV $\gamma$~ray from $^{21}{\rm Ne}(p,\gamma)$ and the 1951~keV $\gamma$ ray $^{22}{\rm Ne}(p,\gamma)$ at $E_p = 1113$~keV.
}
\begin{ruledtabular}
\begin{tabular}{ccccc}
&\multicolumn{2}{c}{HPGe1 ($0^\circ$)}&\multicolumn{2}{c}{HPGe2 ($119^\circ$)}\\
\cline{2-3}\cline{4-5}\\[-0.9em]
Reaction&Energy & FWHM & Energy & FWHM \\
&(keV)&(keV) & (keV) & (keV)\\
\colrule\\[-0.9em]
$^{21}{\rm Ne}(p,\gamma)$&1955.65(10)&2.82(10)&1950.08(10)&3.27(10)\\
$^{22}{\rm Ne}(p,\gamma)$&1951.48(10)&3.83(10)&1950.21(10)&3.15(10)\\
\end{tabular}
\end{ruledtabular}
\end{table}
To better understand the ramifications of this systematic effect we performed Monte Carlo simulations of Doppler effects for these $\gamma$~rays. In these simulations, once the detector geometries were taken into consideration, the lifetime of the decaying state was used to randomly generate decay times from an exponentially distributed probability density function. The energy loss by the recoiling excited nucleus prior to photon emission was calculated using an interpolation routine with tabulated stopping powers from SRIM~\cite{TRIM}. Finally the recoil momentum of the nucleus was folded with the intrinsic detector resolution to obtain the Doppler shifts (and broadenings) for each detector. The Doppler shifted energies and FWHM's of the registered $\gamma$~rays for both the detectors are listed in Table~\ref{tab:doppler}. It is clear that the 1951~keV $\gamma$~ray has a relatively small shift due to the long lifetime of the 2391~keV state in $^{23}$Na~($\tau \approx 600$~fs)~\cite{nndc}. Coupled with the 
relatively 
large Doppler 
broadening of the peak of interest at $\theta_{\gamma} = 119^\circ$, this makes distinguishing between the two peaks futile at this angle. Thus we were compelled to not use the data from HPGe2 at the higher energy proton resonance. On the other hand, the relatively larger separation of the peak centroids at $E_p = 1113$~keV for $\theta_\gamma = 0^\circ$ made it possible to fit the two peaks as shown in Fig.~\ref{fig:fit_to_1952}. On generating coincidences by gating on the $7800 \to 1952$~keV transition in the NaI detector, it is gratifying to obtain a relatively clean coincidence spectrum (\textit{c.f}~Fig.~\ref{fig:coincs}) for this resonance, with no obvious traces of contamination affecting the peaks of interest.  
\begin{table*}[htb]
\caption{\label{tab:results}%
Relative branches obtained for the three $\gamma$ ray transitions of interest from the 1952~keV state. 
}
\begin{ruledtabular}
\begin{tabular}{c.....}
&\multicolumn{4}{c}{Branching fraction (\%)}\\
$E_\gamma$&\multicolumn{2}{c}{$E_p = 908$~keV}&\multicolumn{1}{c}{$E_p = 1113$~keV}&\multicolumn{1}{c}{Adopted\footnote{Obtained from a weighted mean. 
Systematic uncertainties in the evaluated branches have been added in quadrature to the statistical uncertainties.
}}&\multicolumn{1}{c}{Previous}\\
\cline{2-3}\cline{4-4}\\[-0.9em]
(keV)&\multicolumn{1}{c}{HPGe1}&\multicolumn{1}{c}{HPGe2}&\multicolumn{1}{c}{HPGe1}&\multicolumn{1}{c}{value}&\multicolumn{1}{c}{work}\\
\colrule\\[-0.9em]
1295&0.278(69)&0.280(208)&0.256(43)&0.26(5)&0.29(5)\footnote{From Ref.~\cite{nndc}.}\\
1369&99.291(86)&99.202(259)&99.293(57)&99.29(9)&99.70(10)\footnote{From Ref.~\cite{gorres:82}.}\\
1952&0.431(52)&0.519(156)&0.450(38)&0.45(8)&0.61(24)\footnote{From Ref.~\cite{firestone_report}.}\\
\end{tabular}
\end{ruledtabular}
\end{table*}

\subsection{Simulation geometry}
Since the $\gamma$-ray detection efficiencies for this experiment were determined from PENELOPE simulations, potential systematic uncertainties to the efficiencies arise from inaccuracies in the simulation model. To better understand these effects we performed several simulations with conservative estimates of uncertainties in detector distance, detector orientation, source distribution and lead thickness. The differences in the extracted efficiencies (which were of the order of 1\%) were added in quadrature to the statistical uncertainties obtained from simulations using the original model shown in Fig.~\ref{fig:setup}.   

\begin{figure}[b]
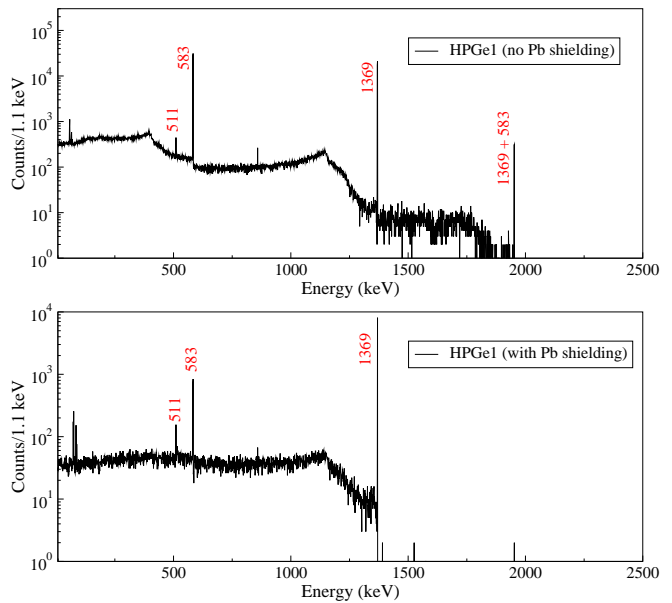

\includegraphics[scale=0.36]{summing_bare_for_paper.eps}\\
\includegraphics[scale=0.36]{summing_total_for_paper.eps}
\caption{\label{fig:summing}(Color online) Monte Carlo simulations for 4~million $1952 \to 583 \to 0$~keV cascades. Top panel: Histogram for an unshielded Ge detector at $\theta_\gamma = 0^\circ$ where the efficiencies show $\sim$~5\% summing corrections. Bottom panel: Simulated result obtained with Pb shielding incorporated as shown in Fig.~\ref{fig:setup}.}
\end{figure}
\subsection{Summing corrections}
The large solid angle subtended by the HPGe detectors made it important to estimate the effects of summing of the $1952 \to 583 \to 0$~keV cascades in the detectors. Such an effect would result in the loss of counts from the 1369~keV peak and in the case of photo-peak summing would result in spurious counts in the 1952~keV peak. It was anticipated that the 2.54~cm~thick lead shielding placed at the front of the detectors would reduce such summing significantly. 
To better understand the summing effects we performed additional Monte Carlo simulations in which we incorporated the two-step cascade mentioned above with a vanishing $1952 \to 0$~keV branch. A comparison of the results both with and without the lead shields is shown in Fig.~\ref{fig:summing}. 
These simulations 
confirm negligible summing corrections for both $\gamma$~rays.  
\section{Results and discussion}

The measured branches of the three $\gamma$~rays of interest are listed in Table~\ref{tab:results}. 
With our value for the \mbox{$1952 \to 0$~keV} branch and the lifetime of the 1952-keV level, \mbox{$\tau = 11.5(2.9)$}~fs~\cite{nndc}, we obtain a partial width of 
\begin{equation}
\label{eq:width}
\Gamma_\gamma = 2.57(79) \times 10^{-4}~{\rm eV}~. 
\end{equation}
This value is more precise but not in disagreement with the result reported in Refs.~\cite{firestone_report,Bowers:99}. Making the same assumptions as Ref.~\cite{Bowers:99}, namely, that the width in Eq.~\eqref{eq:weak} is the partial width of the $1952 \to 0$~keV transition and that the relative signs of $b$ and $c_1$ are as predicted by the shell-model calculation, we obtain 
\begin{equation}
\label{weakmag}
b/Ac_1 = -8.7(1.1)~.
\end{equation}   
Inserting this value in Eq.~\eqref{eq:tensor} yields 
\begin{equation}
d/Ac_1 = 21(6)~,
\end{equation}
which remains significantly larger than expectations. 

We note that the width used in Eq.~\eqref{eq:weak} should be the {\em isovector part of the $M1$ width}. Thus, one needs to determine the fraction of the measured width that corresponds to the isovector $M1$ matrix element.
In the long wavelength limit the $M1$ operator is given by~\cite{deshalit}:
\begin{equation} 
{\bm \mu} \approx \mu_N \sum_i \left [\frac{1+\tau_3(i)}{2} {\bm l}_i + \left\{ 0.88+4.7~\tau_3(i) \right\} {\bm s}_i\right].
\end{equation}   
Because of the large coefficient multiplying the isovector spin part of the operator, $M1$ transitions are usually dominated by their isovector component and the implicit 
hypothesis of Ref.~\cite{Bowers:99} is well justified.
However, in the particular case of $^{22}{\rm Na}$, the matrix element for the spin operator in the analog $\beta$ decay is suppressed. It follows from isospin symmetry and the CVC hypothesis
that the $M1$ matrix element could have a significant isoscalar contribution. However, since the assignments for the states in question are $T=0$ for the $3^+$ state and $T=1$ for the $2^+$ state, this scenario would require further a suppression of the isovector part of the ${\bm l}$ operator. This seems unlikely. The isospin assignments mentioned above are validated by a shell model calculation using the NushellX code with isospin non-conserving interactions~\cite{brown}. 

Next, we consider what fraction of the measured width could be due to the $E2$ component.
On using the USDBcdpn interaction~\cite{richter}, the shell model calculation predicts the width of the transition to be dominated by its $M1$ component, with an $E2/M1$ mixing ratio $\delta \sim 0.02$. 
However,
the $^{22}{\rm Na}$ nucleus is known to have a large deformation ($\beta \sim 0.5$), with well established rotational bands~\cite{Warburton:68,olness:70,garrettNPA:71,garrett:71,macarthur}. 
The $2_1^+$ state was identified as the (collective) rotational excitation of the $0^+$ state at 657~keV. Thus,  
the $2_1^+ \to 3_1^+$ transition is a $\Delta K = 3$ transition, which, in the complete absence of coupling between the collective and intrinsic degrees of freedom would be dominated by the $M3$ multipolarity~\cite{Hamilton}. 
More realistically, the $M1$ and $E2$ multipolarities are not completely forbidden, but considerably hindered and their transition strengths are characterized by a {\em degree of $K$-forbiddenness}~\cite{Hamilton} 
\begin{eqnarray}
\nu = \mid \Delta K \mid -\lambda = 
\left\{ 
\begin{array}{ll}
2 & {\rm for}~M1\\
1 & {\rm for}~E2 
\end{array}
\right.
\end{eqnarray}
where $\lambda$ is the multipolarity of the transition. Empirically, 
this implies that the $M1$ strength could be two orders of magnitude more hindered than the $E2$ component~\cite{Hamilton}. This is at odds with the shell model prediction, but not unexpected, considering that collective excitations are not naturally incorporated in the shell model.
Thus, it is likely that the $M1$ component of the transition is much smaller than the one obtained from the branch.
Assuming a vanishing isovector $M1$ component and thereby setting $b = 0$, we obtain
 \begin{equation}
 d/Ac_1 = 12(6)~,
 \end{equation}
which is roughly consistent with expectations. An alternative scenario is that the $\gamma$ transition is $M1$ dominated, but the relative signs of $b$ and $c_1$ are opposite to that obtained by the shell model. In that case one obtains
\begin{equation}
d/Ac_1 = 3(6)~. 
\end{equation}

\section{Conclusions}
In conclusion, this experiment makes the first unambiguous measurement of the $\Delta T = 1$~$2_1^+ \to 3_1^+$ $\gamma$-ray branch in $^{22}$Na. Assuming that the relative sign of $b$ with respect to $c_1$ is as predicted by the shell model~\cite{Firestone:78} and that the width for the transition is dominated by its $M1$ isovector component, on using the previously measured $\beta-\gamma$ correlation coefficient we obtain an unexpectedly large induced-tensor form factor for $^{22}$Na $\beta$ decay. 
One possible resolution
is that the relative signs of $b$ and $c_1$ are opposite to the theoretical predictions of Ref.~\cite{Firestone:78}. 
Further analysis, taking into account evidence for high deformation, indicates that a more plausible resolution of the dilemma is that the transition is dominated by its $E2$ component. An experiment that determined the $E2/M1$ mixing ratio for the analog transition would resolve this issue.

\begin{acknowledgments}
We are thankful to Gerald Garvey for useful comments, the CENPA staff at UW for help with the accelerator operations, Paul Vetter for providing us a copy of Ref.~\cite{firestone_report} and John Sharpey-Schafer for directing us to Ref.~\cite{macarthur}. 
This work was partially supported by the National Research Foundation of South Africa, the US Department of Energy, the Natural Science and Engineering Research Council of Canada, and the National Research Council of Canada. LP thanks the NRF MANUS/MATSCI program at the UWC for financial support. The UW researchers were supported by the US Department of Energy, Office of Nuclear Physics, under contract number DE-FG02-97ER41020.
CW acknowledges support from the U.S. National Science Foundation under Grant
No. PHY-1102511 and the U.S. Department of Energy, Office of Science, under award No. DE-SC0016052. 
\end{acknowledgments}


\bibliography{ne21pg}

\begin{thebibliography}{48}
\bibitem [{imi()}]{imix}%
  \BibitemOpen
  \href@noop {} {}\bibinfo {note} {Unlike in Ref.~\cite{TH:15},
  isospin-symmetry breaking corrections in $^{22}$Na beta decay have been
  neglected here as they are small compared other uncertainties. We also do not
  assign an uncertainty to $c_1$ as it has a negligible effect.}\BibitemShut
  {Stop}%
\bibitem [{cal()}]{calcs}%
  \BibitemOpen
  \href@noop {} {}\bibinfo {note} {The authors of Ref.~\cite{Firestone:78}
  calculate the suppressed $c_1$ form factor to be $c_1 = 0.00266$. This is
  approximately 6 times smaller than the experimentally extracted value in
  Eq.~\eqref{gt}. The higher-order matrix elements are not similarly
  suppressed. So it is not expected that their calculations would show
  similarly large deviations.}\BibitemShut {Stop}%
\end{thebibliography}

\end{document}